\documentclass[sigconf]{acmart}

\usepackage{makecell}
\usepackage{enumitem}

\definecolor{cML}{RGB}{100,160,200}
\definecolor{cAR}{RGB}{205,120,60}
\definecolor{cDS}{RGB}{70,165,145}
\definecolor{cCIM}{RGB}{240,185,50}
\definecolor{c1}{HTML}{B85450}
\definecolor{c2}{HTML}{006EAF}
\definecolor{c3}{HTML}{7A7A00}
\definecolor{mycirclepurple}{HTML}{694784}

\newcommand{\cmark}{\textbf{\checkmark}}

\newcommand{\xmark}{\ensuremath{\times}}
\newcommand{\myparagraph}[1]{\noindent\textit{\textbf{#1.}~}}

\colorlet{lvlA}{teal!42}
\colorlet{lvlB}{teal!24}
\colorlet{lvlC}{teal!12}
\colorlet{lvlD}{teal!4}
\colorlet{lvlX}{red!18}

\newcommand{\fcirc}[1]{%
  \begingroup
  \setlength{\unitlength}{1ex}%
  \raisebox{-0.65ex}{%
    \begin{picture}(2.25,2.25)%
      \put(1.125,1.125){\color{mycirclepurple}\circle*{2.25}}%
      \put(1.125,1.125){\makebox(0,0){\textcolor{white}{\scriptsize\bfseries #1}}}%
    \end{picture}%
  }%
  \endgroup
}

\AtBeginDocument{%
}

\copyrightyear{2027}
\acmYear{2027}
\setcopyright{cc}
\setcctype{by}
\acmConference[ASPDAC '27]{32nd Asia and South Pacific Design Automation Conference}{January 25--28, 2027}{Tokyo, Japan}
\acmBooktitle{32nd Asia and South Pacific Design Automation Conference (ASPDAC '27), January 25--28, 2027, Tokyo, Japan}
\acmDOI{10.1145/3840404.3848237}
\acmISBN{979-8-4007-2923-2/2027/01}

\begin{document}

\title{IMPACT: Constraint-Aware Search for Imperfectly Factorized Accelerator Mappings}

% ============================================================
% ACM AUTHOR METADATA
% ============================================================

\author{Konstantinos Stavrakakis}
\authornote{These authors contributed equally to this work.}
\affiliation{%
  \institution{Delft University of Technology}
  \city{Delft}
  \country{Netherlands}
}
%\correspondingauthor
\email{kstavrakakis@tudelft.nl}

\author{Ilgaz Er}
\authornotemark[1]
\affiliation{%
  \institution{Delft University of Technology}
  \city{Delft}
  \country{Netherlands}
}
\email{I.Er@student.tudelft.nl}

\author{Theofilos Spyrou}
\affiliation{%
  \institution{Delft University of Technology}
  \city{Delft}
  \country{Netherlands}
}
\affiliation{%
  \institution{Sorbonne Université}
  \city{Paris}
  \country{France}
}
\email{T.Spyrou@tudelft.nl}

\author{Dimitrios Stavrakakis}
\affiliation{%
  \institution{Technical University of Munich}
  \city{Munich}
  \country{Germany}
}
\email{dimitrios.stavrakakis@tum.de}

\author{Stephan Wong}
\affiliation{%
  \institution{Delft University of Technology}
  \city{Delft}
  \country{Netherlands}
}
\email{J.S.S.M.Wong@tudelft.nl}

\author{Georgi Gaydadjiev}
\affiliation{%
  \institution{Delft University of Technology}
  \city{Delft}
  \country{Netherlands}
}
\email{G.N.Gaydadjiev@tudelft.nl}

\author{Said Hamdioui}
\affiliation{%
  \institution{Delft University of Technology}
  \city{Delft}
  \country{Netherlands}
}
\email{S.Hamdioui@tudelft.nl}

\renewcommand{\shortauthors}{Stavrakakis et al.}
\settopmatter{authorsperrow=4}

\begin{abstract}

\looseness=-2
The efficiency of Deep Neural Network (DNN) accelerators depends on how each tensor operation is mapped onto the hardware. A mapping specifies how the computation is tiled across the memory hierarchy, ordered, and parallelized across the processing elements. These choices form a combinatorially large search space, where efficient mappings are hard to find. Most violate hardware constraints, and most remaining mappings underutilize the hardware. Existing mappers either search only part of this space to keep the search tractable, excluding the most efficient mappings, or sample the full space randomly, so most mappings they evaluate are infeasible. We present IMPACT, a mapper that searches the full space. IMPACT pairs a genetic algorithm with a constraint-aware repair operator that converts each infeasible mapping into a feasible one instead of discarding it. On a flexible accelerator, IMPACT outperforms state-of-the-art mappers in EDP, latency, and energy on \textit{every} DeepBench workload. On a compute-in-memory (CIM) accelerator, whose constraints leave few mappings feasible, it reduces EDP by 66\%, latency by 47\%, and energy by 18\% over the strongest prior mapper across five networks, and rivals hand-crafted expert mappings. Across flexible and highly constrained accelerators, IMPACT finds efficient mappings automatically. IMPACT will be released as open source.

\end{abstract}




\maketitle

%\input{Tex/Revisions/Introduction}
%\vspace{-1.5em}
\section{Introduction}

\looseness=-1
DNN workloads are dominated by tensor operations, such as convolutions and matrix multiplications, that demand extreme compute throughput and memory bandwidth \cite{efficient-processing,compute-trends}.
 %Dedicated 
DNN accelerators meet these demands with arrays of processing elements (PEs) that compute in parallel, and a hierarchy of 
%on-chip 
memory buffers that keep data close to the PEs to maximize reuse \cite{tpu,eyeriss}. These buffers, however, hold only a small fraction of data at a time, so the computation must be split into tiles that fit each memory level and distributed across the PEs. Executing an operation therefore requires assigning the computation to the hardware: how large a tile each level holds for each dimension of the operation (\textit{tiling factors}), how many PEs process each dimension in parallel (\textit{spatial unrolling}), the order each level steps through its tiles (\textit{loop order}), and which tensors each level retains (\textit{keep/bypass}). This assignment is called a mapping. 
Once the hardware is fixed, the mapping determines how effectively its resources are used.
Two mappings of the same operation can produce very different memory traffic and PE utilization, and thus may differ by orders of magnitude in latency and energy\cite{timeloop,ruby}.

\begin{figure*}[t]
    \centering
    \includegraphics[width=1\textwidth]{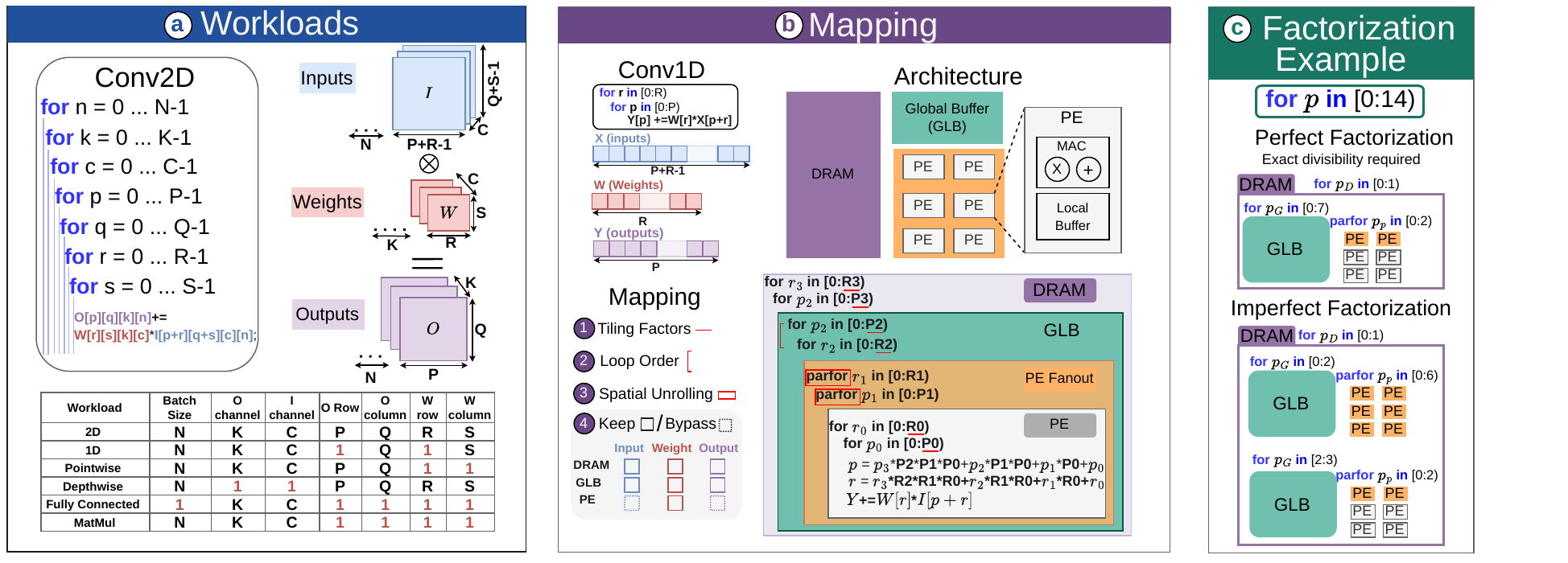}
    \vspace{-2.2em}
    \caption{
    \looseness=-1
    (a) A 2D convolution as a 7-dimensional loop nest, with common layers as special cases. (b) Mapping onto a DRAM–GLB–PE hierarchy through tiling factors, loop order, spatial unrolling, and keep/bypass. (c) Perfect versus imperfect factorization for D=14 on 6 PEs.}
    \label{fig:background-fig}
    \vspace{-1.5em}
\end{figure*}

\looseness=-1
Selecting the best mapping is a combinatorial optimization problem. Because every choice is made per dimension and memory level, the number of possible mappings grows exponentially with both; for a single convolution on a realistic accelerator, the search space contains at least $10^{20}$ candidates \cite{timeloop}. Most of these candidates are infeasible, because every choice must simultaneously satisfy every hardware constraint, such as buffer capacities and PE array dimensions. 
The feasible mappings, therefore, are only a small part of this larger space, yet too large to enumerate, and they are not equally efficient.
%The feasible mappings, therefore, form a smaller space, yet too large to enumerate, and they are not equally efficient.
A spatial unrolling that occupies only part of the PE array leaves the remaining PEs idle, and tiling factors that capture little reuse force the same data to be re-fetched, inflating latency and energy \cite{dataflow-el,data-movement}. 
%Efficient mappings are therefore a small fraction of this feasible subspace, which is itself only a small part of the full space.
Efficient mappings are therefore a small fraction of the already small feasible subspace.
As a result, an effective mapper must satisfy three requirements. It must search the entire space, isolate the feasible subspace, and identify the most efficient mappings within it.

\looseness=-1
Existing mappers meet these requirements only in part. Timeloop \cite{timeloop} and ZigZag \cite{zigzag} restrict the space to tiling factors and spatial unrollings that \textit{perfectly} divide the tensor dimensions, and randomly sample the reduced search space. SALSA \cite{salsa} restricts it further, fixing the spatial unrolling as input and annealing over loop orderings and tiling factors.
This reduced space, however, excludes the most efficient mappings: when the tensor dimensions and the PE array are misaligned, the mappings that fill the array are exactly those excluded, and they remain out of reach however long the reduced space is searched. GAMMA \cite{gamma} searches the same reduced space with a genetic algorithm, but its operators recombine mappings without regard for hardware constraints, so it spends most evaluations on infeasible offspring. Ruby \cite{ruby} expands the space, extending Timeloop's model with tiles and spatial unrollings that need not divide the tensor dimensions, recovering utilization the other mappers sacrifice. However, Ruby samples the expanded space randomly, with no mechanism to only search within the feasible subspace, so nearly all its evaluations are spent on mappings that are infeasible. 
In short, the most efficient mappings are either excluded to keep the search tractable, or admitted into a space that no existing search can navigate. No existing mapper searches the entire space, keeps its evaluations feasible, and finds the most efficient mappings within it.
%that violate hardware constraints. 

\looseness=-1
To address this, we propose 
%We present 
IMPACT, a mapping framework that searches the full mapping space while keeping every candidate feasible, through a constraint-aware repair operator that steers infeasible mappings toward feasible ones. Our key contributions are:
\begin{itemize}[leftmargin=*,topsep=2pt, itemsep=2pt, parsep=1pt, partopsep=1pt]
    \looseness=-1
    \item \textbf{Higher mapping quality:} IMPACT's genetic search finds more efficient mappings than prior automated mappers. On Eyeriss~\cite{eyeriss}, a flexible accelerator, it outperforms both Timeloop and Ruby on \textit{every} DeepBench~\cite{deepbench} workload and efficiency metric.
    \item  \textbf{Search under tight constraints:}
    IMPACT's repair operator keeps the search inside the feasible region even where few mappings are valid. On a CIM-based accelerator, it lowers EDP by 66\%, latency by 47\%, and energy by 18\% compared to Timeloop across five networks, and rivals expert hand-crafted mappings.
    \looseness=-1
    \item \textbf{Efficient search:} 
    %By steering infeasible candidates to nearby feasible mappings instead of discarding them, IMPACT spends its budget on feasible mappings instead of wasting it. On Eyeriss, 83\% of attempts yield a feasible mapping, against Timeloop's 10\% and Ruby's 5\%.
    By steering infeasible candidates to nearby feasible mappings instead of discarding them, IMPACT spends its budget on mappings that can execute. On Eyeriss, 83\% of attempts yield a feasible mapping, against Timeloop's 10\% and Ruby's 5\%.

\end{itemize}

\looseness=-1
The remainder of this paper is organized as follows. Section \ref{sec:background} introduces the mapspace and feasibility constraints, and Section \ref{sec:problem} formulates the constrained search problem. Section \ref{sec:framework} presents IMPACT, and Section~\ref{sec:repair} details its repair operator. Section~\ref{sec:evaluation} evaluates IMPACT against state-of-the-art mappers. Sections \ref{sec:discussion} and \ref{sec:conclusion} discuss implications and conclude.

\vspace{-1.1em}
\section{Background}
\label{sec:background}
\vspace{-0.5em}
%\myparagraph{Mapping Search Space}A tensor layer maps directly to a loop nest in which each tensor dimension defines one loop bound and each loop-body iteration performs one MAC~\cite{timeloop}. \autoref{fig:background-fig}a depicts a 2D convolution expressed as a 7-dimensional loop nest, alongside a table that presents how common layer types reduce to the same nest by fixing subsets of the dimensions to one. Fig~\autoref{fig:background-fig}b depicts an example mapping of the nest onto a two-level memory hierarchy. A mapping specifies four decisions (\autoref{fig:background-fig}b): the tiling factors per dimension per level, the loop ordering within each level, the spatial unrolling of selected loops across PEs, and the bypass decisions per tensor per level. The product of tiling factors for a given dimension across all levels must equal the workload dimension; this product constraint defines the tiling subspace. The set of all mappings consistent with a hardware specification is the mapspace; its size grows combinatorially with the number of loop dimensions and memory levels~\cite{timeloop}.

\looseness=-1
\myparagraph{Mappings}%
Fig.~\ref{fig:background-fig}b illustrates such a mapping, using a simpler 1D convolution with dimensions $R$ and $P$ on a DRAM–Global Buffer(GLB)–PE hierarchy, and marks the four decisions a mapping specifies. For every dimension and every memory level, a \textit{tiling factor} (\fcirc{1}) sets the size of the tile that level holds. Within each level, the \textit{loop order} (\fcirc{2}) fixes the sequence in which its tiles are traversed. At the PE level, the \textit{spatial unrolling} (\fcirc{3}) selects the dimensions whose tiles execute in parallel across the array, up to its fanout. For every tensor and every level, a \textit{keep/bypass} (\fcirc{4}) decision sets whether that level buffers the tensor. The four decisions are coupled through the tiling factors. The product of a dimension's factors across all levels must equal the dimension itself. This \textit{product constraint} ties the choices of every memory level together. The set of all mappings consistent with a hardware target is the mapspace, and it grows combinatorially with the number of dimensions and memory levels\cite{timeloop}.

\looseness=2
\myparagraph{Imperfect factorization}Standard mappers satisfy the product constraint with tiling factors that divide each dimension exactly, a condition called perfect factorization\cite{timeloop,zigzag}. When no divisor of a dimension matches the PE count, some PEs remain idle in every iteration. Fig.~\ref{fig:background-fig}c illustrates this for a dimension D=14 on 6 PEs. The closest divisor is 2, leaving 4 of the 6 PEs idle throughout all iterations. Imperfect factorization relaxes the divisibility requirement by allowing a remainder tile \cite{ruby}. The main tile is sized to fill the PE array, and only the final iteration executes a smaller remainder tile. In the same example, a tile size of 6 occupies all PEs for two iterations and 2 PEs for the last, reducing idle cycles. This improves PE utilization, but enlarges the mapspace, since tile sizes are no longer restricted to divisors of $D$.

\begin{figure*}[t]
    \centering
    \includegraphics[width=0.9\linewidth]{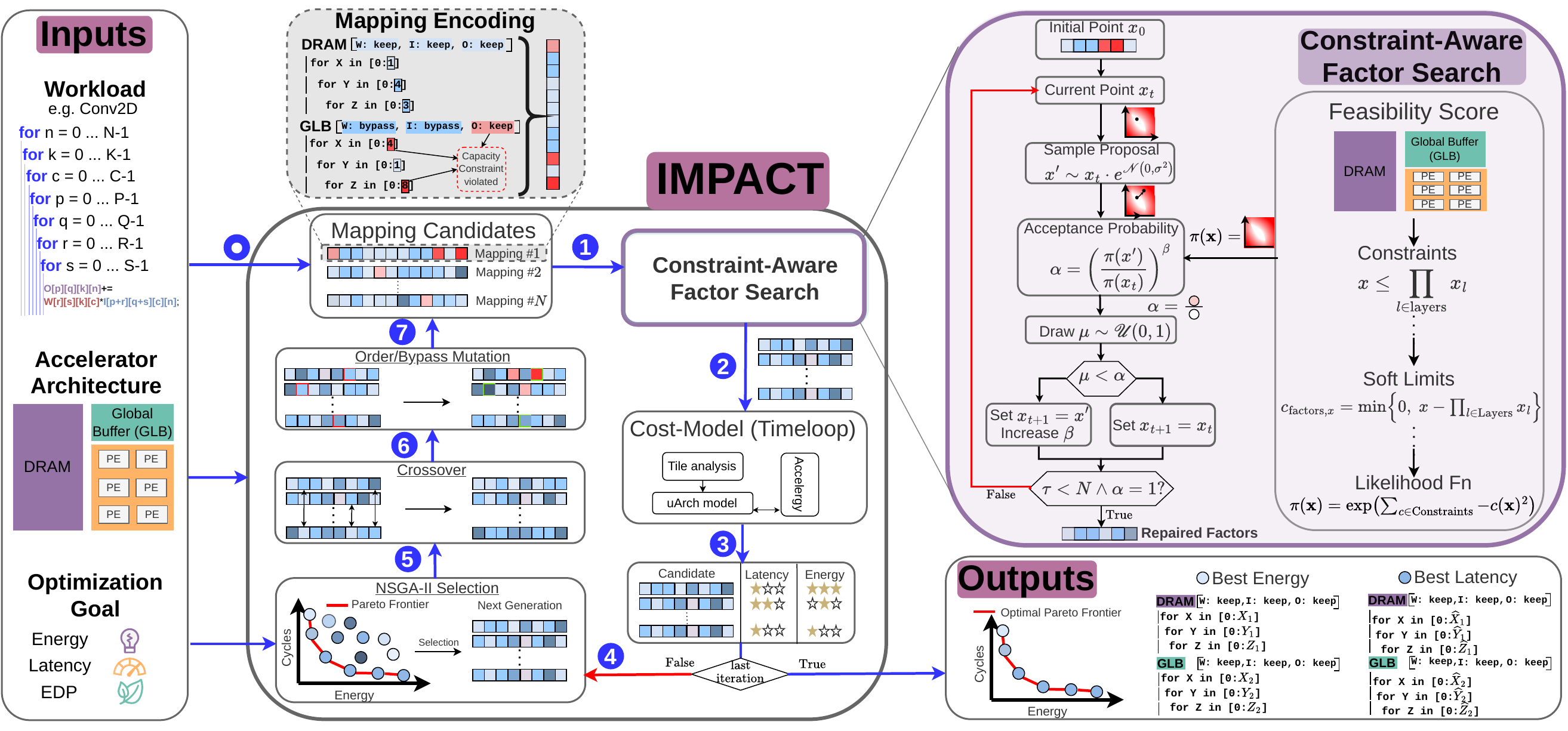}
    \vspace{-1.5em}
    \caption{IMPACT framework overview. A genetic algorithm and an annealed Metropolis-like repair operator map a DNN workload and accelerator specification to a Pareto-optimal mapping set, evaluated by Timeloop.}
    \label{fig:framework_overview}
    \vspace{-1.5em}
\end{figure*}

%Standard mappers require each tiling factor to be an exact divisor of the workload dimension, a condition called perfect factorization \cite{timeloop} or prime factorization \cite{zigzag}. When no exact divisor matches the available PE count, some PEs remain idle every iteration. Fig.~\ref{fig:background-fig}c illustrates this for dimension D=14 with 6 PEs: the closest perfect factor is 2, leaving 4 PEs idle throughout all iterations. Imperfect factorization relaxes the divisibility constraint by allowing a remainder tile \cite{ruby}. The main tile is sized to fill the PE array; only the final iteration uses a smaller partial tile. In the same example, a tile size of 6 utilizes all PEs for two full iterations and 2 PEs for the last, reducing idle cycles. This improves PE utilization at the cost of enlarging the mapspace, since remainder sizes are no longer restricted to divisors of $D$.

\myparagraph{Feasibility Constraints}Imperfect factorization expands the map-space with candidates that may leave loop iterations incomplete, exceed PE fanout, or overflow memory capacity \cite{ruby}. Feasibility constraints identify the subset of this expanded space that is valid and executable \cite{timeloop}. \textit{Workload coverage} constraints require that all loop iterations complete and bound any dummy work introduced by remainder padding. \textit{Spatial fanout} constraints require that the spatially unrolled tile fits within the available PE array and interconnect fanout. \textit{Storage capacity} constraints require that the tensor tiles resident at each memory level fit within its physical capacity; bypassed tensors do not consume capacity at that level and are excluded from this check. 
%We refer to a mapping that satisfies all three constraints as feasible or executable.
A mapping that satisfies all three constraints is feasible.
The feasible mapspace is the search domain for any mapping framework operating under imperfect factorization.

\myparagraph{Workloads}Several DNN layers map directly to a loop nest: each dimension of the layer defines one loop, and each iteration of the loop body performs one multiply-accumulate (MAC)\cite{timeloop}. Fig.~\ref{fig:background-fig}a shows a 2D convolution expressed as a seven-dimensional loop nest. As the accompanying table shows, common layer types, such as matrix multiplications and depthwise convolutions, reduce to the same nest with some dimensions fixed to one. A single loop-nest form therefore covers all these workloads. Executing one on an accelerator requires mapping its loops to the hardware.

\vspace{-1.5em}
\section{The Mapping Optimization Problem}
\vspace{-0.5em}
\label{sec:problem}

\myparagraph{Problem Formulation}
A hardware target $H$ is defined by its memory hierarchy, PE array dimensions, buffer capacities, and fanout limits at each level. A workload $W$ is defined by its per-dimension loop bounds. A mapping $M$ is a tuple of tiling factors per dimension per level, loop order per level, spatial unrolling, and keep/bypass decisions per tensor per level. The feasible mapspace $\mathcal{F}(W, H)$ is the set of all $M$ that simultaneously satisfy workload coverage, spatial fanout, and storage capacity constraints. The mapping problem is to solve:
{
\setlength{\abovedisplayskip}{1pt}
\setlength{\belowdisplayskip}{1pt}
\setlength{\abovedisplayshortskip}{1pt}
\setlength{\belowdisplayshortskip}{1pt}
\[
M^* = \underset{M \in \mathcal{F}(W,H)}
{\operatorname{argmin}}\, C(M,W,H)
\]
}
\noindent where $C(M, W, H)$ is a cost function over energy, latency, or a weighted combination of both. The decision variables are discrete, the constraints couple them non-locally across memory levels, and $C$ is non-convex over $M$. Enumerating $\mathcal{F}(W, H)$ exhaustively is therefore intractable at realistic problem scales.

\looseness=-1
\myparagraph{Search Complexity}Finding an efficient mapping is hard for two compounding reasons. First, the feasibility constraints must hold simultaneously and couple decisions non-locally. A tiling factor at one level determines tile sizes at all levels below, jointly determining whether storage and fanout limits are met \cite{timeloop}. The feasible mapspace $\mathcal{F}(W, H)$ therefore occupies a small, irregularly shaped fraction of the full expanded space. Second, within $\mathcal{F}(W, H)$, cost is highly non-uniform. Two feasible mappings can differ by orders of magnitude in energy and latency \cite{timeloop}. The high-quality region is consequently a small fraction of $\mathcal{F}(W, H)$.
%already-small feasible space. 
Exhaustive search is intractable and random sampling wastes most evaluations on infeasible or dominated candidates. Finding $M^*$, or a close approximation, requires a search strategy that can navigate a large, sparsely feasible, and highly non-uniform space.

\vspace{-1em}
\section{Optimization Framework}
\label{sec:framework}
\vspace{-0.5em}
\subsection{Architecture and Design Rationale}
\vspace{-0.3em}
Section~\ref{sec:problem} characterized the mapping problem as a search over a large, discrete space that is sparsely feasible and non-uniform in cost. Efficient mappings are rare and scattered, and the space offers no gradient to follow, so IMPACT searches it globally with a genetic algorithm  that explores broadly while concentrating on the strongest mappings it finds. A genetic algorithm alone cannot keep that search feasible. Loop orders 
%and keep/bypass flags 
are independent choices, so recombining them preserves feasibility. The tiling factors of a dimension and the keep/bypass flags, however, are coupled through the storage capacity constraint, and any change at one level can invalidate the factorization as a whole. IMPACT therefore repairs the factors of each mapping with a dedicated local operator before it is scored. The framework has three components:

%A genetic algorithm alone cannot keep that search feasible: loop order and bypass are discrete choices genetic operators recombine well, whereas tiling factors and spatial unrolling are coupled non-locally across memory levels, and naive recombination cannot hold them feasible. IMPACT therefore repairs the factors of each mapping with a dedicated local operator before it is scored. The framework has three components:

\begin{itemize}[
    leftmargin=1.2em,
    labelsep=0.35em,
    itemsep=0pt,
    topsep=0pt,
    parsep=0pt,
    partopsep=0pt
]
%[leftmargin=*]
    \item A genetic algorithm evolves the global structure of each mapping across generations.
    %\item A constraint-aware operator expresses the factor space as a likelihood function and uses Metropolis-Hastings~\cite{metropolis-hastings} with simulated annealing~\cite{simulated-annealing} to repair the tiling factors of each offspring into a feasible configuration.
    \item A constraint-aware operator expresses constraint violation as a feasibility score and uses a Metropolis-style acceptance rule~\cite{metropolis-hastings,metropolis1953equation} with simulated annealing~\cite{simulated-annealing} to repair the tiling factors of each offspring toward a feasible configuration.
    \item Timeloop evaluates each mapping analytically and returns energy, latency, and EDP \cite{accelergy}.
\end{itemize}
Fig.~\ref{fig:framework_overview} shows the pipeline, which maps a DNN workload and a hardware specification to a Pareto-optimal set of mappings. IMPACT samples an initial population across the mapspace without enforcing validity, since the operator repairs every offspring before it is scored. Each generation applies crossover and mutation, repairs the tiling factors of the offspring, evaluates the repaired mappings with Timeloop, and selects the next generation with NSGA-II \cite{nsga2}. Since energy and latency trade off and no single mapping minimizes both, NSGA-II keeps a population of individuals close to the Pareto frontier. The loop repeats until the evaluation budget is spent.

\vspace{-1.3em}
\subsection{Representation and Genetic Operators}
\vspace{-0.4em}
%\looseness=-1
\myparagraph{Mapping Encoding}IMPACT encodes each candidate as a chromosome that captures the complete mapping. A mapping is a loop nest, the same form that the analytical model evaluates, so each gene corresponds directly to a loop bound, a loop order, or a buffering choice in the evaluated mapping. At each memory level, each workload dimension maps to a loop whose bound sets the tile size at that level, and at the PE level, selected dimensions are unrolled spatially across the array. The chromosome carries three gene classes: loop-order genes fix the nesting order at each level, factor genes set the loop bounds at each level, and keep/bypass genes select which tensors each level buffers. Each gene also records the hardware restrictions on its decision, such as a level that cannot buffer a given tensor, so the operators never violate these restrictions during search.

\looseness=-1
\myparagraph{Crossover}Crossover recombines and perpetuates the structural traits that high-performing mappings share. Keep/bypass and loop-order genes recombine independently under a uniform scheme, making no structural assumption about gene cross-correlation \cite{uniform-crossover}. The tiling factors of a dimension are interdependent through the product constraint, so they recombine as a single unit that preserves the dimension's factorization.

\myparagraph{Mutation}Mutation introduces the variation that recombination alone cannot reach, keeping the search from stalling. It handles the discrete genes directly, permuting a loop order or flipping a keep/bypass flag. The legal values of a dimension's tiling factors, however, form a thin, irregular set that a random perturbation almost always misses, so mutation cannot explore them, and the offspring it produces are almost all infeasible.
%Mutation introduces the variation that recombination alone cannot reach, keeping the search from stalling. It handles the discrete genes directly: for loop orders and bypass flags, it permutes an order or flips a flag. The tiling factors are harder. Their legal values form a thin, irregular set that a random perturbation almost always misses, so mutation cannot explore them, and the offspring it produces are almost all infeasible.

%\vspace{-2.5em}
\section{Constraint-Aware Factor Search}
\label{sec:repair}
%\vspace{-0.5em}
%IMPACT reaches this fragmented feasible set with a repair operator: after crossover and mutation, it moves the infeasible factors of an offspring to a nearby feasible set before the mapping is scored. 

\looseness=-1
After crossover and mutation, most offspring carry infeasible tiling factors. 
IMPACT pairs its genetic search with a repair operator that moves these factors to a nearby feasible set before the mapping is scored.
The feasible set, however, is fragmented into disconnected regions, so moving only through feasible factors cannot cross between them. The operator therefore uses an annealed Metropolis-style walk that steps through infeasible ground between the regions. Fig.~\ref{fig:framework_overview} depicts the walk. Starting from the offspring's factors, each step proposes a new factor set, scores it by its distance from feasibility, and accepts or rejects it, until the walk settles on a feasible set.

\looseness=-1
\myparagraph{Proposal Distribution}Given the factors of an offspring after crossover and mutation, each step proposes a new factor set by scaling the current factors. IMPACT draws each proposal from a log-normal centered on the current factors and rounds to the nearest positive integer, since factors are integer-valued. Because the factors combine multiplicatively and range over orders of magnitude, the proposal scales them rather than adding to them; the log-normal is symmetric in log-space, so an up- or down-scaling by the same ratio is equally likely.

\myparagraph{Constraint Model}To compare a proposal of factors with the current factors, the operator scores each by its distance from feasibility, so it can rank two infeasible sets and move toward the feasible region. We write $f_{l,d}$ for the factor at level $l \in L$ and dimension $d \in D$, and define feasibility by three constraints:

\begin{enumerate}[
    leftmargin=1.2em,
    labelsep=0.35em,
    itemsep=0pt,
    topsep=2pt,
    parsep=0pt,
    partopsep=0pt
]
%[leftmargin=*]
    \item \textbf{Workload coverage.} The factors of a dimension must cover the workload dimension $F_d$, with a bounded remainder under imperfect factorization:
    \begin{equation*}
    \setlength{\abovedisplayskip}{1pt}
    \setlength{\belowdisplayskip}{1pt}
    \setlength{\abovedisplayshortskip}{1pt}
    \setlength{\belowdisplayshortskip}{1pt}
    F_d \leq \prod_{l \in L} f_{l,d} \leq F_{d,\max}
    \end{equation*}

    \item \textbf{Spatial fanout.} The factors at a spatial level $l$ must fit the $N_l$ hardware components 
    %and occupy at least half of them: 
    \begin{equation*}
    \setlength{\abovedisplayskip}{0pt}
    \setlength{\belowdisplayskip}{0pt}
    \setlength{\abovedisplayshortskip}{0pt}
    \setlength{\belowdisplayshortskip}{0pt}
    %0.5\,N_l \leq
    \prod_{d \in D} f_{l,d} \leq N_l
    \end{equation*}

    \item \textbf{Storage capacity.} The resident tensor tiles at a memory level $l$ must fit within its capacity $C_l$:
    \begin{equation*}
    \setlength{\abovedisplayskip}{0pt}
    \setlength{\belowdisplayskip}{0pt}
    \setlength{\abovedisplayshortskip}{0pt}
    \setlength{\belowdisplayshortskip}{0pt}
    S_l \leq C_l
    \end{equation*}
\end{enumerate}

\noindent Hard limits would give the search no direction \cite{constraint-handling}, so the operator turns each violation into a penalty, such that an infeasible factor set carries a graded score. The feasibility score:
\begin{equation*}
\setlength{\abovedisplayskip}{0pt}
\setlength{\belowdisplayskip}{0pt}
\setlength{\abovedisplayshortskip}{0pt}
\setlength{\belowdisplayshortskip}{0pt}
\pi(\mathbf{x}) = \exp\!\Big(-\!\!\sum_{c \in \mathcal{C}} c(\mathbf{x})^2\Big)
\end{equation*} 
sums the squared margins, peaking at feasibility and decaying smoothly as the total violation grows.

\begin{figure}[t]
    \centering
    %\vspace{-0.8em}
    \includegraphics[width=0.85\linewidth]{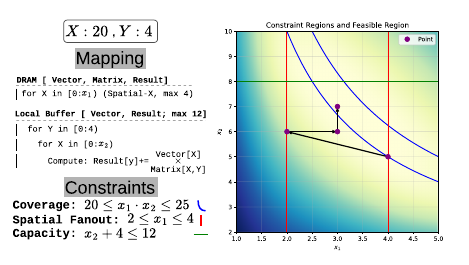}
    \vspace{-2.1em}
    %\addtocounter{figure}{-2} % Go back to Figure 3
    \caption{Repair walk toward the feasible region}
    \vspace{0em}
    \label{fig:random-walk}
\end{figure}

\begin{figure*}[t]
    \centering
    
    \includegraphics[width=0.98\textwidth]{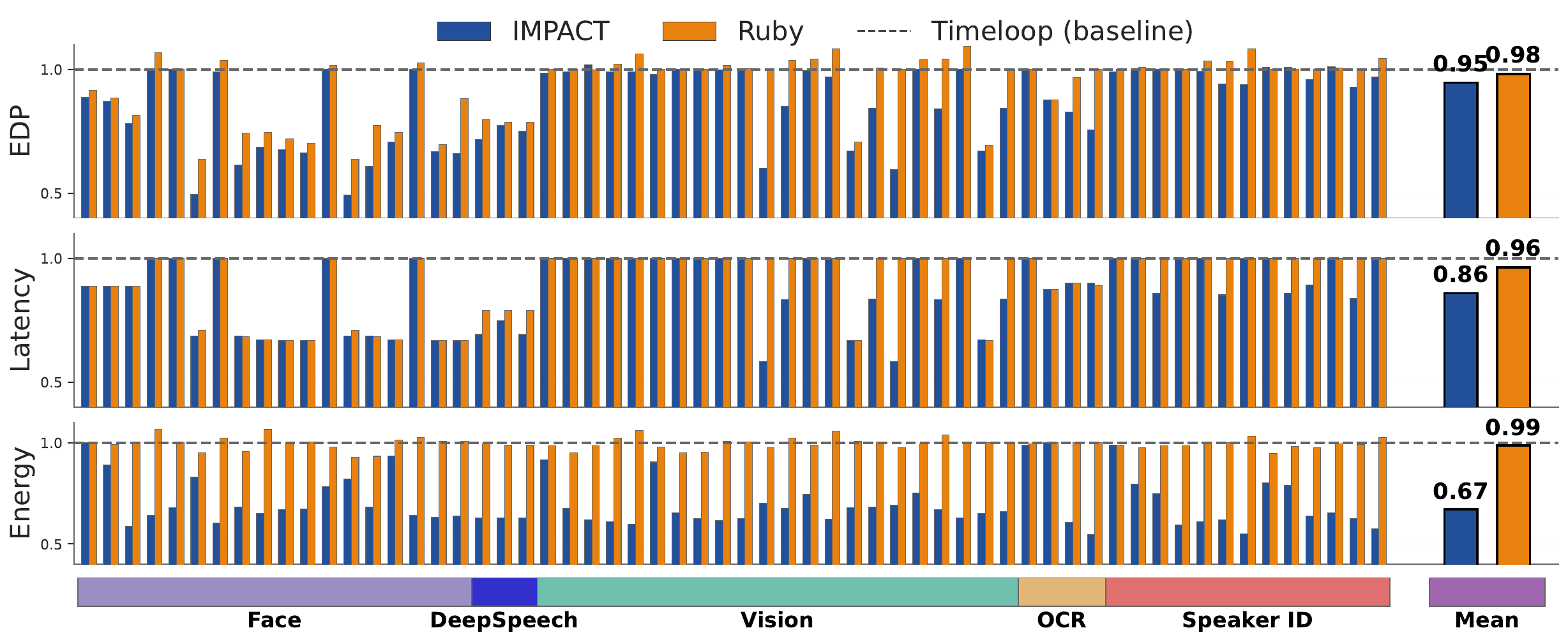}
    \vspace{-1.5em}
    %\addtocounter{figure}{1}
    \caption{Energy, latency, and EDP on Eyeriss across DeepBench workloads for IMPACT and Ruby, normalized to Timeloop.}
    %\vspace{-1em}
    \label{fig:eyeriss-comparison-deepbench}
\end{figure*}

\looseness=-1
\myparagraph{Acceptance}This soft score lets the operator repair factors with a Metropolis-style acceptance rule. The operator always accepts a proposal that lowers the violation, and accepts one that raises it with a probability  equal to the score ratio of the proposal to the current set. Only the final set need be feasible, so the operator may accept steps through infeasible ground, crossing between the disconnected regions that a feasibility-preserving move could not. Because the feasibility score falls exponentially with violation, large moves into infeasibility are rarely accepted. 
%Thus, the search stays mostly near the feasible regions and reaches one without stalling where several constraints meet.
The walk therefore stays near the feasible regions and reaches a feasible set even where several constraints are tight.
%Thus the search stays near the feasible regions and reaches one even where several constraints meet, without stalling.

\looseness=-1
\myparagraph{Annealing and Termination}A fixed acceptance rule would either converge prematurely or fail to converge, so a simulated annealing schedule at temperature $1/\beta$ sharpens it. Exploration runs at high temperature for a fixed number of steps, then $\beta$ grows and steepens the penalties, funneling the steps onto the feasible region. The walk stops at the first feasible set it reaches. Local minima in the constraint space can still trap it, so the sharpened phase is bounded: a walk that exhausts its steps restarts from the offspring's factors. This way, the repair stays near the factor choices the search is optimizing.

Figure~\ref{fig:random-walk} shows the complete operator on the two factors $x_1$ and $x_2$ of a single dimension. The vertical lines bound $x_1$ by fanout, the horizontal line bounds $x_2$ by storage, and the curved band bounds their product by coverage; together they leave only the narrow feasible sliver where all three hold. From a feasible start, the operator proposes an infeasible step, the penalty scores it, and the acceptance rule carries the walk toward the feasible region. The walk moves through infeasible ground rather than navigating the thin sliver, since only the final set must be feasible. It stops at the first feasible point, a few steps from its start. Applied to every offspring, this repair makes the imperfect-factorization search tractable.

%\vspace{-1.2em}
\section{Evaluation}
\label{sec:evaluation}
\vspace{-0.2em}

%\vspace{-0.5em}
\looseness=-1
The evaluation tests whether (i) IMPACT finds better mappings than prior automated mappers, and (ii) whether it sustains that quality on hardware whose constraints leave few mappings valid. We compare against Timeloop\cite{timeloop}, which defines the mapspace and cost model, and Ruby\cite{ruby}, the only prior mapper that searches the imperfect-factorization space. GAMMA\cite{gamma}, ZigZag\cite{zigzag}, and SALSA\cite{salsa} are built on different cost models that only evaluate perfectly factorized mappings, and cannot score the mappings this work targets.
The first experiment evaluates the mappers on Eyeriss, a flexible architecture on which prior mappers already perform well, and isolates the quality of the search itself. The second moves to a compute-in-memory architecture, where a deep memory hierarchy and rigid tiling rules make feasible mappings rare, and adds expert hand-crafted mappings to test whether IMPACT remains effective where constraint-unaware search breaks down. 
The third measures each method's valid-mapping yield, characterizing how each search navigates the feasible region under the same time constraint.
%The third measures the rate at which each method finds valid mappings, characterizing how each search navigates the feasible region under the same time constraint.

\begin{comment}
\begin{table}[h]
%\vspace{0.5em}
\centering
\caption{IMPACT search configuration.}
\label{tab:hyperparameters}
\vspace{-1.2em}
\footnotesize

\begin{tabular}{@{}c@{\hspace{5em}}c@{}}
\toprule
\textbf{Search} & \textbf{Repair} \\
\cmidrule(r{0.5em}){1-2}
\cmidrule(l{0.5em}){1-2}

\begin{tabular}[c]{@{}l@{\hspace{4em}}r@{}}
Population size    & $100$ \\[0.45\baselineskip]
Initial candidates & $400$ \\[0.45\baselineskip]
Elite survivors    & $50$
\end{tabular}
&
\begin{tabular}[c]{@{}l@{\hspace{1.2em}}r@{}}
Proposal width $\sigma$  & $0.15$ \\
Inverse temp.\ $\beta_0$ & $1$ \\
Cooling factor           & $1.$3 \\
Exploration / max steps  & $3$ / $5000$
\end{tabular}
\\
\bottomrule
\end{tabular}
\vspace{-1.8em}
\end{table}
\end{comment}

\begin{table}[h]
\centering
\caption{IMPACT search configuration.}
\label{tab:hyperparameters}
\vspace{-1.3em}
\footnotesize

\begingroup
\setlength{\aboverulesep}{0.2ex}
\setlength{\belowrulesep}{0.2ex}

\begin{tabular}{@{}c@{\hspace{5em}}c@{}}
\toprule
\textbf{Search} & \textbf{Repair} \\
\midrule

\begin{tabular}[c]{@{}l@{\hspace{4em}}r@{}}
Population size    & $100$ \\[0.45\baselineskip]
Initial candidates & $400$ \\[0.45\baselineskip]
Elite survivors    & $50$
\end{tabular}
&
\begin{tabular}[c]{@{}l@{\hspace{1.2em}}r@{}}
Proposal width $\sigma$  & $0.15$ \\
Inverse temp.\ $\beta_0$ & $1$ \\
Cooling factor           & $1.3$ \\
Exploration / max steps  & $3$ / $5000$
\end{tabular}
\\[-0.6ex]
\bottomrule
\end{tabular}

\endgroup
\vspace{-2.2em}
\end{table}

%\vspace{-1.8em}
\subsection{DeepBench Evaluation on Eyeriss}
%\vspace{-0.5em}
\label{subsec:deepbench-eyeriss}
\looseness=-1
We compare IMPACT against Timeloop and Ruby on Eyeriss, evaluating on the DeepBench suite used in Ruby for direct comparison. 
%Each mapper searches every layer for the same wall-clock time and retains the best mapping found for each objective. 
Each mapper runs 32 times per layer with independent seeds under the same per-run wall-clock budget of 15 minutes, 
%and we report the best mapping found across runs for each objective.
and we report the mean across runs for \textit{each} objective, so the results do not depend on a favorable seed.
All three use the same analytical cost model, so any difference in mapping quality reflects the search alone. Energy, latency, and EDP are normalized to the Timeloop baseline. IMPACT uses the same search configuration throughout, listed in Table~\ref{tab:hyperparameters}.

\looseness=-1
\myparagraph{Comparison to automated mappers}
IMPACT outperforms both Timeloop and Ruby on every workload family and every objective. As shown in Figure~\ref{fig:eyeriss-comparison-deepbench}, the largest reduction is in energy. IMPACT lowers it by 33\% over Timeloop at the mean across workloads, while Ruby lowers it by 1\%. Latency falls by 14\%, and EDP by a more modest 5\%.

\looseness=-1
\myparagraph{Effect of workload shape}IMPACT's gains for EDP and latency concentrate on layers where perfect factorization cannot fill the PE array.
%IMPACT's gains for EDP and latency concentrate on layers whose dimensions perfect factorization cannot map onto the PE array without leaving it underused.
%IMPACT's gains for EDP and latency concentrate on layers that perfect factorization cannot fit to the PE array. 
Vision layers whose dimensions align with Eyeriss's 14x12 PE array map well under Timeloop, so Ruby and IMPACT add little. On Face and DeepSpeech, filling the array requires imperfect factorization. Here Ruby reduces latency below Timeloop, and IMPACT reduces it further. %IMPACT's largest gains are in energy. Imperfect factorization lets it size tiles for reuse that perfect factorization misses, so its mappings re-fetch less data and spend less energy.
For minimum-energy mappings, IMPACT strongly outperforms both. By uniquely exploring the imperfectly factorized tilings space, it finds mappings with lower data movement.

\looseness=-1
\myparagraph{Source of the gains} IMPACT's advantage on Eyeriss comes from the structure of its search. Eyeriss is a flexible architecture, and its loose constraints leave many mappings valid, so Timeloop and Ruby already sample good ones. IMPACT searches the wider valid space and covers it. The repair operator keeps its evaluations feasible across the imperfectly factorized region, and the genetic algorithm concentrates those evaluations around the strongest mappings it finds. Random sampling stops at the good mappings; this coverage is what finds the better ones.

\begin{figure}[t]
    \centering
    \includegraphics[width=0.95\linewidth]{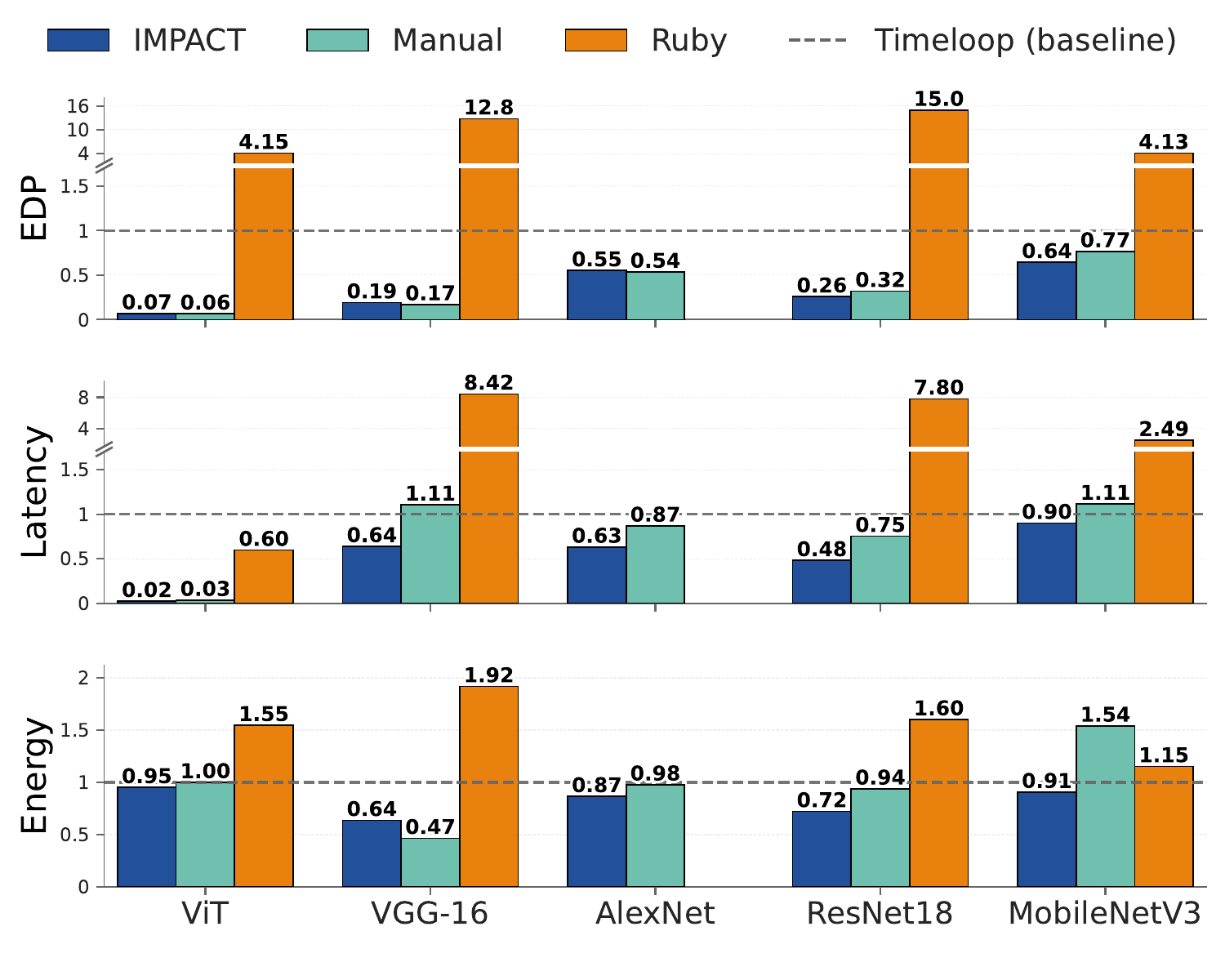}
    \vspace{-2em}
    %\addtocounter{figure}{+1}
    \caption{Energy, latency, and EDP across five networks on a CIM architecture, normalized to Timeloop.}
    %Energy, latency, and EDP on a CIM-architecture across five networks for IMPACT, expert (Manual), and Ruby, normalized to Timeloop.}
    \vspace{-2.5em}
    \label{fig:cim-arch-comparison}
    
\end{figure}

%\vspace{-1em}
\subsection{Evaluation on CIM Architecture}
\label{subsec:cim-evaluation}
\vspace{-0.5em}

%We next evaluate IMPACT on a compute-in-memory (CIM) architecture whose dataflow is far more constrained than the Eyeriss spatial array. The architecture is based on the analog RRAM crossbar of \cite{wan2022compute} and modeled in CimLoop\cite{cimloop}. Its fixed reduction dataflow imposes rigid tiling rules, so feasible mappings are sparse and most candidates are invalid. 
\looseness=-2
We next evaluate IMPACT on a compute-in-memory architecture whose dataflow is far more constrained than the Eyeriss spatial array. We use a provided CimLoop\cite{cimloop} architecture, based on the analog RRAM crossbar of \cite{wan2022compute}, whose fixed reduction dataflow is representative of CimLoop's CIM models. This dataflow imposes rigid tiling rules, so feasible mappings are sparse and most candidates are infeasible.
We compare IMPACT against Timeloop and Ruby on five networks: a Vision Transformer (ViT)\cite{ViT}, VGG-16\cite{vgg}, AlexNet\cite{alexnet}, ResNet-18\cite{resnet}, and MobileNetV3\cite{mobilenetv3}. For this architecture we additionally compare against the hand-crafted mappings provided by CimLoop. Each mapper searches every layer for the same wall-clock time, following the same 32-run protocol as Section~\ref{subsec:deepbench-eyeriss}, reporting the mean across runs.
%following the same protocol as Section~\ref{subsec:deepbench-eyeriss}.
%Each mapper searches every layer for the same wall-clock time. 
Network energy and latency are summed across layers, while network EDP is computed as their product.
All metrics are normalized to the Timeloop baseline.

\myparagraph{Comparison to automated mappers}
%IMPACT outperforms the hand-crafted expert mappings on most configurations, without any manual tuning. It improves on the experts in 11 of the 15 metric-workload cases. The expert mappings are tuned to minimize EDP, and they improve on IMPACT in only four cases, three of them in EDP, with an average discrepancy of 4.2\% and a maximum of 17\% on VGG-16. 
IMPACT is the only automated mapper that improves on Timeloop across every network and every metric. As Figure ~\ref{fig:cim-arch-comparison} shows, averaged over the five workloads, it reduces EDP by 66\%, latency by 46.6\%, and energy by 18.2\% relative to Timeloop. Ruby searches a larger space than Timeloop but a smaller one than IMPACT; even so, without a structured search it reaches no high-quality mappings. On this architecture it underperforms even Timeloop, inflating EDP by $4\times$ to $15\times$ and failing for several AlexNet layers.

\myparagraph{Comparison to hand-crafted mappings}IMPACT outperforms the hand-crafted expert mappings on most configurations, without any manual tuning. It improves on the experts in 11 of the 15 metric-workload cases. The expert mappings are tuned to minimize EDP, and they improve on IMPACT in only four cases, three in EDP and one in energy, with an average discrepancy of 4.2\% and a maximum of 17\% on VGG-16. IMPACT reaches the lowest latency on all five networks and lower energy on four of the five, improving on the experts by 32\% and 20\% on average, respectively. IMPACT therefore matches the experts on the objective they are tuned for, and outperforms them on the objectives they are not.

\myparagraph{Key takeaway}IMPACT's advantage over prior mappers widens as the architecture becomes more constrained. On a flexible architecture, loose constraints leave many mappings valid, so every mapper finds competitive solutions and IMPACT's margin is modest. On a constrained architecture, Timeloop is confined to perfect factorization and cannot fill the array, while Ruby searches an expanded space without structure and mostly evaluates invalid or inefficient candidates. IMPACT's constraint-aware repair lets the genetic algorithm keep exploring and exploiting the sparse feasible region throughout the search. On constrained hardware, this reaches the quality that until now required hand-tuning.
%This is what lets it reach, on constrained hardware, the quality that until now required hand-tuning.
%This is what allows it to outperform both prior mappers and hand-crafted mappings, positioning IMPACT as a design-automation tool that generates expert-quality mappings.

\begin{comment}
    
\begin{figure*}[t]
    \centering
    %\vspace{-0.8em}
    \includegraphics[width=1\textwidth]{Fig/valid_yield_bar.pdf}
    %\vspace{-1.5em}
    \caption{
    \looseness=-1
    Valid-mapping yield on Eyeriss across DeepBench workloads for Ruby, Timeloop, GA (without repair), and IMPACT.}
    \label{fig:valid_yield}
\end{figure*}
\end{comment}

\vspace{0.5em}
\begin{figure}[h]
    \centering
    \includegraphics[width=1\linewidth]{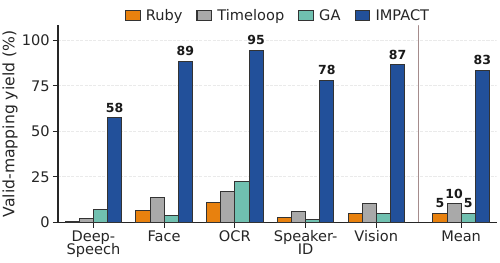}
    \vspace{-2em}
    \caption{Valid-mapping yield on Eyeriss across DeepBench workloads for Ruby, Timeloop, GA (without repair), and IMPACT.}
    \vspace{-2em}
    \label{fig:valid_yield}
\end{figure}

\vspace{0.5em}
\subsection{Valid-Mapping Yield}
\label{subsec:feasible-sample-rate}
%\vspace{-0.5em}
%\vspace{-0.3em}
%We characterize each search by its feasible-sample rate on Eyeriss. Alongside the three mappers, we measure a plain genetic algorithm, obtained by removing the repair operator from IMPACT, to isolate the repair operator's contribution to the yield. The rate is the fraction of candidates that satisfy all workload-coverage, spatial-fanout, and storage-capacity constraints. We report it per DeepBench workload, under the same wall-clock budget as in Section~\ref{subsec:deepbench-eyeriss}.
\looseness=-1
We characterize each search by its feasible-sample rate on Eyeriss. 
Alongside the three mappers, we measure a plain genetic algorithm, obtained by removing the repair operator from IMPACT. Like GAMMA, it searches without constraint awareness, and isolates the repair operator's contribution to the yield.
%Alongside the three mappers, we measure a plain genetic algorithm, obtained by removing the repair operator from IMPACT, to isolate the repair operator's contribution to the yield. 
The yield is the fraction of candidates that produce a mapping satisfying all workload-coverage, spatial-fanout, and storage-capacity constraints. A candidate is one attempt at a feasible mapping: a sampled mapping for Timeloop and Ruby, an offspring for the plain GA, and one repair walk for IMPACT, whose yield is therefore the fraction of walks that reach a feasible set rather than restart. We report it per DeepBench workload, under the same wall-clock budget as in Section~\ref{subsec:deepbench-eyeriss}.

\myparagraph{Valid-mapping yield}IMPACT spends its budget on feasible mappings, while the baselines spend most of theirs on infeasible ones. As Figure~\ref{fig:valid_yield} shows, it reaches 83\%, against 10\% for Timeloop, 5\% for Ruby, and 5\% for the plain genetic algorithm. The yields separate the search strategies. Ruby samples the imperfect-factorization space without checking feasibility, and 95\% of its candidates are invalid. 
The plain GA reaches only 5\%. Its recombination barely preserves feasible structure, so most offspring violate constraints and are discarded. A genetic algorithm without repair produces no more feasible mappings than random sampling. The repair operator lifts the yield to 83\% by steering each offspring to the feasible set, and is what makes search possible in this space.
%The plain GA reaches 5\%: its recombination barely preserves feasible structure, but most offspring violate constraints and are evaluated only to be discarded. The repair operator lifts the yield to 83\% by steering each offspring toward the feasible set, and is what makes the search viable in this space.

\looseness=-1
\myparagraph{Key Takeaway}The repair operator's feasibility yield explains the quality gains of the previous experiments. In Sections~\ref{subsec:deepbench-eyeriss} and~\ref{subsec:cim-evaluation}, IMPACT ran under the same budget as the baselines, yet found better mappings. With its budget spent on feasible candidates rather than invalid ones, IMPACT's search covers far more of the real mapspace in the same time, and finds the better mappings within it.

%\vspace{-1.3em}

\newpage

\section{Discussion}
\label{sec:discussion}
\vspace{-0.5em}
\myparagraph{Comparison to prior work}Table~\ref{tab:framework_comparison} shows that no prior mapper handles both difficulties of mapping search: imperfect factorization and hardware constraints. Timeloop\cite{timeloop}, ZigZag\cite{zigzag}, GAMMA\cite{gamma}, and SALSA\cite{salsa} all restrict search to perfectly-factorized mappings; SALSA further fixes spatial unrolling as an input. GAMMA's genetic operators do not enforce dataflow constraints, confining its search to fully flexible accelerators. Ruby\cite{ruby} comes closest, opening the imperfect-factorization space that holds the best mappings, but searches without structure or constraint awareness, so most candidates are invalid or low-quality. Unlike these, IMPACT pairs a genetic algorithm with a constraint-aware repair operator, addressing both imperfect factorization and hardware constraints, and reaches the high-quality mappings the others miss.

\myparagraph{Practical impact}IMPACT's ability to find high-quality mappings for a given tensor accelerator enables systematic design-space exploration across the software-hardware stack. Hardware engineers can measure how a change in the architecture affects performance, without manually searching for the best mapping at each configuration. Compiler engineers can retarget a workload to a new accelerator with an IMPACT-generated mapping, instead of re-tuning the mapping for every architecture by hand.

%\myparagraph{Limitations and Future Work}IMPACT relies on an analytical cost model and inherits its accuracy. Because the search is independent of this model, a higher-fidelity evaluator such as RTL or silicon measurement can replace it without changing the search. A natural next step is to integrate IMPACT into a compilation flow that generates code from its mappings.
\vspace{-1.2em}
\begin{table}[t]
\centering
\caption{
\vspace{-0.2em}
\looseness=-1
Comparison between state-of-the-art mappers
%Comparison of search frameworks for mapping optimization.
}
\vspace{-1.3em}
\label{tab:framework_comparison}
\renewcommand{\arraystretch}{1.2}
\resizebox{\columnwidth}{!}{%
\begin{tabular}{c c c c c c c c}
\toprule[2pt]
\textbf{Framework} &
\textbf{Strategy} &
\makecell{\textbf{Imperfect}\\\textbf{Factorization}} &
\makecell{\textbf{Loop}\\\textbf{Ordering}} &
\makecell{\textbf{Tile}\\\textbf{Size}} &
\makecell{\textbf{Spatial}\\\textbf{Unrolling}} &
\makecell{\textbf{Dataflow}\\\textbf{Configurability}} &
\makecell{\textbf{Constraint}\\\textbf{Awareness}} \\
\midrule[2pt]

Timeloop\cite{timeloop} &
Random &
\xmark &
\cmark &
\cmark &
\cmark &
\cmark &
\xmark \\

Ruby\cite{ruby} &
Random &
Spatial &
\cmark &
\cmark &
\cmark &
\cmark &
\xmark \\

GAMMA\cite{gamma} &
GA &
\xmark &
\cmark &
\cmark &
\cmark &
Flexible Only &
\xmark \\

ZigZag\cite{zigzag} &
Random &
\xmark &
\cmark &
\cmark &
\cmark &
\cmark &
\xmark \\

SALSA\cite{salsa} &
SA+Random &
\xmark &
\cmark &
\cmark &
\xmark &
\cmark &
\xmark \\

\textbf{IMPACT} &
\textbf{GA+Repair} &
\textbf{Full} &
\textbf{\cmark} &
\textbf{\cmark} &
\textbf{\cmark} &
\textbf{\cmark} &
\textbf{\cmark} \\

\bottomrule[2pt]
\end{tabular}
}
\vspace{-2.5em}
\end{table}

%\vspace{-0.5em}
\section{Conclusion}
\label{sec:conclusion}
\vspace{-0.7em}
\looseness=-1
In this work, we present IMPACT, a framework for finding high-quality mappings on tensor accelerators, including highly constrained ones. It uses a genetic algorithm for structured search of the imperfect-factorization space, and a repair operator that keeps the search feasible as constraints tighten. IMPACT consistently outperforms prior automated mappers, and on constrained accelerators it rivals hand-crafted expert mappings. In practice, IMPACT decouples automated mapper performance from how constrained the hardware is. It therefore extends automated mapping to specialized accelerators, which until now depended on expert tuning.

\vspace{-1.2em}
\begin{acks}
\vspace{-0.5em}
This work was supported by the European Union's Horizon
Europe programme through Ferro4EdgeAI (Grant Agreement
No.~101135656). The views expressed are solely those of
the authors and do not necessarily reflect those of the
European Union or the granting authority, neither of
which can be held responsible for them.

\end{acks}

\vspace{-1em}
\bibliographystyle{ACM-Reference-Format}
\bibliography{bib/riaibib}

\end{document}